\documentclass[12pt]{article}
\usepackage{amsmath,amsfonts,amssymb,latexsym}
\usepackage{epsfig}
\usepackage{graphicx}

\newcommand{\sect}[1]{\setcounter{equation}{0}\section{#1}}

\def\I{\mathbb{I}}
\def\N{\mathbb{N}}
\def\R{\mathbb{R}}

\begin{document}

\begin{center}
\bf{\LARGE A constructive presentation of rigged Hilbert spaces
\footnote{Contribution to 7th Int. Workshop DICE2014,
Sept. 15-19, 2014, Castiglioncello (Italy), see {\it J.Phys.:Conf.Series}
}
}
\end{center}

\bigskip\bigskip

\begin{center}
Enrico Celeghini
\end{center}

\begin{center}
{Dipartimento di Fisica, Universit\`a di Firenze, \\ I50019 Sesto Fiorentino, Firenze, Italy}
\medskip

{e-mail:celeghini@fi.infn.it}
\end{center}

\begin{abstract}

We construct a rigged Hilbert space for the square integrable functions on the line $L^2(\R)$ adding to the generators of the Weyl-Heisenberg algebra
a new discrete operator, related to the degree of the Hermite polynomials. All together, continuous and discrete operators, constitute the generators of the projective algebra $io(2)$.
$L^2(\R)$ and the vector space of the line $\R$ are shown to be isomorphic representations
of such an algebra and, as both these representations are irreducible, all operators
defined on the rigged Hilbert spaces $L^2(\R)$ or $\R$
are shown to belong to the universal enveloping algebra of $io(2)$.
The procedure can be extended to orthogonal and pseudo-orthogonal spaces of
arbitrary dimension by tensorialization.

Circumventing all formal problems the paper proposes a kind of toy
model, well defined from a mathematical point of view, of  rigged Hilbert spaces
where, in contrast with the Hilbert spaces, operators with different cardinality
are allowed.

\end{abstract}

\sect{Introduction}

Physical quantities we take into account in quantum mechanics are both discrete and continuous.

In the free particle case, position and energy
have both the cardinality $\aleph_1$ but, when we consider localized states, the energy
has a discrete spectrum i.e. the cardinality $\aleph_0$, while position remains continuous.

In a formal description, this causes problems as, in a Hilbert space (HS), dimensions are well defined so that we cannot in the same HS  have operators with a spectrum of
different cardinality.

For this reason the standard approach depicts
localized systems in a countable HS where energy is
diagonalizable while position, that is continuous, is described by means of elaborated limits on functions with compact support.  We are indeed compelled to a complicated and formally unsatisfactory description of position that cannot be described by an operator.

In reality at the beginning of quantum mechanics this problem was not there, as a
physical state was not described by a vector in a HS but as a ray in such a Hilbert space i.e.
it was associated to a whole family of vectors defined up to an arbitrary complex number.
The one-to-one correspondence was thus between a physical state and an element of a
rigged Hilbert space (RHS),
an intricate concept involving a Gelfand triple\;  $\phi \subseteq {\cal H} \subseteq \phi'$ ,
where ${\cal H}$ is a Hilbert space,\;  $\phi$ (dense subset of  ${\cal H}$) is the  $``ket"$ space and\,  $\phi'$ (dual of  $\phi$)\, is the  $``bra"$ space \cite{Bo}. However quickly RHS has been considered an unnecessary complication
as the results can be found, at least at the formal level usually accepted by physicists, in
the Hilbert space obtained representing, by means of the axiom of choice, each entire
ray with a vector of norm one and phase zero.
In concrete the expectation value of the
operator\, $\Omega$\, in the state\, $|\psi\rangle$\, was defined as \cite{Ro}:
\[
\langle \,\Omega\, \rangle\; =\; \frac{\langle \psi | \Omega | \psi \rangle}{\langle \psi | \psi \;\rangle} \; \; ;
\]
where the norm of the vector\, $| \psi \rangle$ is arbitrary and the
operator\, $\Omega$ can relate vectors not necessarily of the same norm.
In the evolution of the theory this freedom has been disregarded and a one-to-one
relation was established between physical states and normalized vectors in a Hilbert space,
preserving only the phase freedom related to the gauge theories.

Furthermore also group theory suggests that
the ``right'' space could be the RHS and not the simpler HS.
For instance in \cite{LiNa} continuous bases are
described jointed to the discrete one into the same
representation of\, $SU(1,1)$ \,and the HS defined by the discrete basis  is
implemented introducing the space
of differentiable vectors and its dual\, $\phi'$ \,that contains the ``generalized eigenvectors'' of the
non compact generators \cite{Ma}.

Moreover by means of rigged Hilbert spaces the Dirac formalism can be reproduced, for the harmonic oscillator,  as shown by B\"ohm in his book \cite{Bo}.

We do not attempt to discuss here the formal properties of RHS , as did B\"ohm \cite{Bo}, but we exhibit a one dimensional example where all formal problems
are circumvented using special functions. Anyway this example is in reality
quite more then an example as it can be easily extended, by tensorial construction, to
orthogonal and pseudo-orthogonal spaces of any dimension.

In conclusion, let us stress that the difference between RHS
formalism and the usual HS one appears to be minor from the physicists
point of view but is essential from the mathematical point of view and leads
to a tremendous mathematical simplification: in fact it justifies the
mathematically undefined operations that the physicists have been
accustomed to in their calculations.

Our fundamental statement is thus that
the restriction from RHS to HS is unjustified in the sense that we lose more of what
we gain in the reduction, because in RHS all is mathematically well defined
and all observables can be correctly described. In addition, as shown in the following,
RHS allow to include inside Lie algebras and Lie universal enveloping algebras
operators with spectrum of different cardinality.

The fundamental ingredients of the paper are well known:

1) Hermite functions with their discrete label and continuous
variable that are described by observables of different cardinality.

2) Lie algebras, groups and representations of special functions.

A more detailed discussion of the algebraic properties of special functions and of
their role as transition matrices between discrete and continuous bases can be
found in \cite{CeOl13}, while in \cite{CeOl14} the technical aspects of the introduction
of RHS by means of
special functions are discussed, showing the solidity of the approach.

Here, on the contrary, our attention is addressed to stress that the RHS is an enough simple
and  mathematically satisfactory formalism.

\sect{Hermite  functions\,   $\{\psi_n(x)\}$\, as localized wave functions on the line}

As a first step, let us introduce the Hermite functions \cite{NIST}
\[
\psi_n(x):=\; \frac {e^{-x^2/2}}{\sqrt{2^n n! \sqrt{\pi}}}\; H_n(x)\, .
\]

As \cite{Fo}
\[
\int_{-\infty}^{\infty} \psi_n(x)\; \psi_{n'}(x)\; dx = \delta_{n,n'}\,\,,\qquad\quad
\sum_{n=0}^\infty\,  \psi_n(x)\;\, \psi_n(x')\, =\; \delta(x-x') \, ,
\]
$\{\psi_n(x)\}$ is  a basis of the space of real/complex square
integrable functions on the line\; $L^2((-\infty, \infty))\, \equiv\, L^2(\R)$ .

Using the freedom of fixing the origin as well as the scale of position and momentum, Hermite functions can be  generalized to  $\psi[n, x_0, s, x]$ so that

\[
\psi[n, x_0, s, x] := \, \frac{e^{-\frac{(x-x_0)^2}{2 s^2}}}     {\sqrt{2^n n! s \sqrt{\pi}}}
H_n \left[(x-x_0)/s\right]
\]

\[
\psi[n, p_0, s, p] :=\, \frac{e^{-\frac {(p-p_0)^2 s^2}{2}} \sqrt{s}}{\sqrt{2^n n!  \sqrt{\pi}}}
H_n \left[(p-p_0)s\right]
\]

\[
\int_{-\infty}^{\infty}  \psi[n, x_0, s, x]\;  \,\psi[ n' ,x_0, s, x]\;  dx \;=\;  \delta_{n n'}
\]

\[
\sum_{n=0}^\infty\,  \psi[n,x_0,s,x] \;  \psi[n,x_0',,s,x']\, =\; s \,\;\delta[(x-x_0)-(x'-x_0')]\, .
\]

Among the $\infty$-many bases of square integrable functions on the
line $L^2(\R)$,\,
Hermite functions are particularly suitable
to connect classical and quantum physics.
Indeed they describe wave packets in the position and, as\,
$\psi_n(x)$ \,is an eigenvector of the Fourier transform,  also in the
momentum; so that they correspond to our intuitive vision of quantum mechanics and, in the appropriate limit, allow to reconstruct
the classical pattern.

Note also that they allow to describe systems with the appropriate behavior in
function of energy. The minimal indetermination corresponds to the state with\, $n=0$ ,\,
while as\, $n$ \,increases, and with\, $n$ \,the energy of the state, also the indetermination increases.

Indeed as

\[
\int_{-\infty}^{\infty}(x-x_0)^2\;  \psi[n, x_0, s, x]^2\; dx \;=\; (n+1/2)\; s^2 \;,
\]
\[
\int_{-\infty}^{\infty}  \psi[n, x_0, s, x]\;\;  \partial_x^2 \,\psi[ n ,x_0, s, x]\;  dx \;=\; -(n+1/2) / s^2 \;,
\]
we have
\[
\Delta X = \sqrt{n+1/2}\; s \,, \qquad  \Delta P = \sqrt{n+1/2} / s
\]
and thus
\[
\Delta X \;\; \Delta P \;\;= n+1/2 \;.
\]

\sect{Algebra of Hermite functions}

The basic idea is  to introduce the operator $N$ that read the label $n$
of the $\{\psi_n(x)\}$ \cite{CeOl13}.
In addition to $X$ and $P \equiv {\bf i} D_x$
we put thus in the space $L^2(\R)$
the operators
$N$ and  ${\I}$ such that
\begin{equation}\label{oper}
 X \psi_n(x) :=  x\, \psi_n(x)\,,\;\;  D_x \psi_n(x) := \psi_n'(x)\,,\;\;
N \psi_n(x) :=  n\, \psi_n(x)\,,\;\; {\mathbb I}\, \psi_n(x) := \psi_n(x) .
\end{equation}

This allows to rewrite the recurrence relations of Hermite polynomials
\[
H'_n(x) - 2\, x\, H_n(x)\, =\, H_{n+1}(x),\qquad  H'_n(x)\, =\, 2\, n\, H_{n-1}(x)
\]
as
\begin{equation}\label{apsi}
a^\dagger\;\, \psi_n(x)\, =\, \sqrt{n + 1}\;\, \psi_{n+1} (x)\,, \;\qquad
a\;\, \psi_n(x)\, =\, \sqrt{n}\;\, \psi_{n-1}(x) ,
\end{equation}
where $a^\dagger$ and $a$ are defined in terms of the
hermitian operators $X$ and $P$
\[
a^\dagger :=\, \frac{1}{\sqrt{2}}\, \left(X-{\bf i} P\right)\,\,\,\,,\qquad
a:=\, \frac{1}{\sqrt{2}}\, \left(X+{\bf i} P\right)\,\,\,.
\]
The algebra contains
the rising and lowering operators on the Hemite functions \cite{CeOl13, Lo}
 \[
[N, a^\dagger] = a^\dagger ,\qquad[N, a] = -a \,\,,\qquad [a,a^\dagger]= \mathbb I \,\,,
\qquad[ \mathbb I ,\bullet] = 0
\]
and is isomorphic to the projective algebra $io(2)$ \cite{Ha, Ba}:
\begin{equation}\label{io2}
[N,X] = -{\bf i} P\,\,,\qquad [N,P] = {\bf i} X\,\,,\qquad [X,P] = {\bf i}\, \I\,\,,\qquad
[ \mathbb I ,\bullet] = 0 \,.
\end{equation}

As discussed in \cite{CeTa}, eqs.(\ref{oper},\ref{apsi}) do not
define the algebra (\ref{io2}) but only one of its
representations.
The Casimir operator
\begin{equation}\label{C}
C\, \equiv\, (X^ 2 - D_x^2)/2 - (N + 1/2)\,\I\, =\,  \{a^\dagger, a \}/2 -(N + 1/2)\,\I
\end{equation}
has indeed zero value on the square integrable functions on the line $L^2(\R)$,
where we can assume $\I=1$ and write
\[
C\, \psi_n(x)  \; =\; \left[ \{a^\dagger , a \}/2 -N - 1/2\,\right] \psi_n(x)\; =\;0
\]
or, alternatively,
\[
C\; \psi_n(x)\; =\; \left[(X^ 2 - D_x^2)/2 - N - 1/2 \right ]\; \psi_n(x)\; =\; 0 \,\,
\]
that, by inspection, are equivalent to the Hermite differential equation:
\[
H''_n(x) - 2\, x\, H'_n(x) + 2\, n\, H_n (x) = 0 \,.
\]
The eq. $C=0$ can be thus also considered as the operatorial identity that defines\, $L^2(\R)$:
\begin{equation}\label{id}
N\; \equiv\; \left(X^ 2 - D_x^2 - 1\right)/2\; \equiv\; \{a,a^\dagger\}/2 -1/2 \,.
\end{equation}

Let us stress that, usually,
the operator $N$  is included inside the UEA of the Weyl-Heisenberg algebra as\,
$N:= a^\dagger a$ \cite{Co},
while here it has been introduced starting from the label of Hermite polynomials and it has
the role of an independent generator of the algebra $io(2)$. Only when the representation $C=0$
is considered (the one of the Hermite functions) the results of the two approachs coincide.

\sect{The line $\R$ and its bases}

 To construct the bases of\, $\R$, \,we move now to group theory \cite{Wu}.

We start from the unitary irreducible representations of the
translation group\; $T^1$
\vspace{.2cm}
\[
P\, |p\rangle\; =\; p\; |p\rangle\,,\qquad \qquad U^p(x)\; |p\rangle\; =\; e^{-{\bf i} p x}\; |p\rangle \, .
\]
The regular representation\; $\{|p\rangle\}$\;  \; ($-\infty \,<\, p \,<\, \infty$) is such that
\[
\langle\, p\,|\,p'\,\rangle\; = \sqrt{2 \pi}\; \delta(p-p')\,,\qquad\qquad
\quad \frac{1}{\sqrt{2 \pi}}\;\int_{-\infty}^{+\infty} |p\rangle\, dp\, \langle p|\, =\, {\mathbb I}
\]

Now we can move, by means of the Fourier transform, to the continuous group element
label $x$ from the irreducible representation label $p$ (also continuous) resorting,
in this way,  the strict connection of group theory with harmonic analysis.

The conjugate basis\, $\{|x\rangle\}$, defined by the operator $X$, is obtained,
indeed, as
\[
|x\rangle:=\left[\frac{1}{\sqrt{2 \pi}}
\int_{-\infty}^{+\infty}\, dp\; e^{-{\bf i} p x}\right]|p\rangle\, ,
\quad
\langle\, x\,|\,x'\,\rangle\; = \sqrt{2 \pi}\; \delta(x - x')\,\,,\quad
 \frac{1}{\sqrt{2 \pi}}\int_{-\infty}^{+\infty}\, |x\rangle\, dx\, \langle x|\,  =\, {\mathbb I}\,\,
\]
and the operators $X$ and $P$ close, together with $\I$, the Weyl-Heisenberg algebra.

Consistently with the previous
section,
we introduce now the set of vectors\; $\{|n\rangle\}$
\begin{equation}\label{defn}
|n\rangle\; :=\; (2 \pi)^{-1/4}\,\int_{-\infty}^{\infty}\, dx\;\, \psi_n(x)\; |x\rangle
\qquad\qquad  n \in \N\; ,
\end{equation}
that, by inspection, is an orthonormal and complete set in\, $\R$
\[
\langle\, n\,|\,n'\,\rangle
= \delta_{n\,n'}\,\,,\qquad\qquad
\sum_{n=0}^{\infty}\; |n\rangle\, \langle n|\;  =\; \mathbb I\,\,  .
\]
$\{|n\rangle\}$\; is thus a discrete basis in the real line\; $\R$\,, i.e. $\R \equiv
\{|p\rangle\} \equiv \{|x\rangle\} \equiv \{|n\rangle\}$
and\, $\{\psi_n(x)\}$ are the transition matrices between\, $\{|n\rangle\}$ \,and\, $\{| x \rangle \}$ :
\[
\langle n | x \rangle \;=\; (2 \pi)^{1/4} \; \psi_n(x) \; .
\]
Relations among the three bases are easily established, as\; $\{\psi_n(x)\}$\, are
eigenvectors of Fourier transform,
\[
\left[\frac{1}{\sqrt{2 \pi}}
\int_{-\infty}^{\infty} dx\; e^{\,{\bf i} p x}\right]\psi_n(x)\;  =\; {\bf i}^n\; \psi_n(p)\,\,,
\]
\[
|\,x\, \rangle\; =\;(2 \pi)^{1/4}\; \sum_{n=0}^\infty \, \psi_n(x)\; |\,n\rangle\,\,, \qquad\qquad
|p\,\rangle =\left[\frac{1}{\sqrt{2 \pi}}
\int_{-\infty}^{\infty}\, dx\, e^{\, {\bf i} p x}\right]\, |x\,\rangle\,\,,
\]
\[
 |n\rangle\; =\; {\bf i}^n\; (2 \pi)^{-1/4}  \,\int_{-\infty}^{\infty}\,
dp\; \psi_n(p) \; |p\rangle\,\,,\qquad\qquad
|\,p\, \rangle\; =\;(2 \pi)^{1/4}\; \sum_{n=0}^\infty  \,{\bf i}^n\; \psi_n(p)\; |\,n \rangle \,.
\]

For an arbitrary vector\; $|f\rangle$ $\in$ $\R$ we thus have
\[
|f\rangle\;\;  =\quad\frac{1}{\sqrt{2\pi}} \int_{-\infty}^{+\infty}\, dx\;
f(x)\; |x\rangle \quad  = \quad
\frac{1}{\sqrt{2\pi}} \int_{-\infty}^{+\infty}\, dp\; f(p)\; |p\rangle \quad  =\quad
\sum_{n=0}^\infty \, f_n\; |n\rangle\,\,\,\,,
\]
\[
f(x) := \langle x|f\rangle\;  =\;(2 \pi)^{1/4} \sum_{n=0}^\infty\, \psi_n(x)\;f_n\,\,, \quad
f(p):= \langle p|f\rangle\;  =\;(2 \pi)^{1/4} \sum_{n=0}^\infty\,(-{\bf i})^n\, \psi_n(p)\;f_n\,\,,
\]
\[
f_n\, :=\; \langle n  | f\rangle\;  =\;\; (2 \pi)^{-1/4} \int_{-\infty}^{+\infty} \,dx\;\psi_n(x)\; f(x)
\;\;  =\;\;\;  {\bf i}^n(2 \pi)^{-1/4}\int_{-\infty}^{+\infty} \,dp\;  \psi_n(p)\, f(p)
\]
and the wave functions\, $f(x)\,, f(p)$ and the sequence\, $\{f_n\}$\; describe
$|f\rangle$ in the three bases.

All seems trivial, but\;  $\{|n\rangle\}$\;  has the cardinality of the natural numbers\;  $\aleph_0$ and, as  all bases in a Hilbert space have the same cardinality,
the structure we have constructed (the quantum space on the line $\R$) is not
a Hilbert space but a rigged Hilbert space, where
\[
|x\rangle \,\in \phi' \,, \quad\qquad |p\rangle \,\in \phi'\, , \quad\qquad |n\rangle \,\in \, \phi \,.
\]

Because of Eq.(\ref{defn}), $\R$ and $L^2(\R)$) are isomorphic and we can write the
algebra\, $io(2)$ \,on $\{|n\rangle\}$ that is, as on\, $\{\psi_n(x)\}$, a
representation  of\, $io(2)$ \,with zero value of the Casimir operator (\ref{C}).

Again on the representations we can assume $\I=1$ and write

\[
X |n\rangle = (a^\dagger + a)  |n\rangle = \sqrt{n+1} |n+1\rangle + \sqrt{n} |n-1\rangle \;,
\]

\[
P  |n\rangle = {\bf i} (a^\dagger - a)  |n\rangle = {\bf i} \sqrt{n+1} |n+1\rangle
- {\bf i} \sqrt{n} |n-1\rangle,
\]

\[
 N |n\rangle = n |n\rangle \;,\qquad \I |n\rangle = |n\rangle \;\qquad \,,
a^\dagger |n\rangle = \sqrt{n+1} |n+1\rangle \;,\;\; a |n\rangle = \sqrt{n} |n-1\rangle \,,
\]

\[
C\, |n\rangle\; =\; \left[
(X^ 2 - D_x^2)/2 - (N + 1/2)\,\right] |n\rangle\; =\;
\left[ \{a, a^\dagger \}/2 -N - 1/2\,\right] |n\rangle\; =\;0 \,,
\]
and the identity  ({\ref{id})
defines not only the vector space\, $L^2(\R)$ but also the vector space $\R$.

Note that, if we restrict ourselves to a Hilbert space, half of the above presented relations are meaningless.

\sect{Universal enveloping algebra and operators in a RHS}

Let us discuss now the implications of the algebraic discussion of previous section
on the operators defined on\, $L^2(\R)$ \,and\, $\R$.

In a RHS   we have no problems to consider
$X$ and $P$
as generators of a Lie algebra together
with the number operator $N$ \cite{Bo,CeOl13}.
This allows us to include differential operators inside the algebraic structure and
to extend the set of operators defined in the universal enveloping algebra.

Both on\, $\{\psi_n(x)\}$ and on\, $\{|n\rangle\}$, the representations are irreducible, so that
-on both spaces $L^2(\R)$\,and \;  $\R$-
all operators of the UEA$[io(2)]$ are defined and
an isomorphism exists between the UEA$[io(2)]$ and the space
of the operators\;  $\{{\cal O}[L^2(\R)]\}$\;  and $\{{\cal O}[\R]\}$ :
\[
 \{{\cal O}[L^2(\R)]\}\, \equiv\, UEA[io(2)] \,\equiv\, \{{\cal O}[\R]\}\,\,,
\]
i.e. each operator\;\; ${\cal O}$\;\; can be written
\[
{\cal O}\; = \sum c_{\alpha\,\beta\,\gamma}\; X^\alpha\, {D_x}^{\beta}\, N^\gamma \;\;
=\;\; \sum d_{\alpha\,\beta\,\gamma}\; {a^\dagger}^\alpha\, N^\beta\, a^\gamma .
\]

From the analytical point of view, an ordered monomial\;\;
$X^\alpha\, {D_x}^{\,\beta}\, N^\gamma  \in$ UEA$[io(2)]$
is an order\, $\beta$\, differential operator but, because of the
operatorial identity (\ref{id}), we have
\[
D_x^2\; \equiv\; X^ 2 - 2 N - 1 ,
\]
and any operator in $\{{\cal O}[L^2(\R)]\}$ and in $\{{\cal O}(\R)\}$ can be written in the form
\[
{\cal O}\, =\, f_0(X)\, g_0(N)\; +\; f_1(X)\; D_x\; g_1(N)
\]
as all higher power of\, $D_x$ \,can be removed and substituted by functions
of\, $X$ \,and\, $N$.
In particular, on the vector\, $\psi_n(x)$\; we have thus
\[
{\cal O}\; \psi_n(x)\; = \; f_0(x)\, g_0(n)\, \psi_n(x)\; +\; f_1(x)\, g_1(n)\; {\psi'}_n(x) \,.
\]

\section{Conclusions}

Rigged Hilbert spaces are shown to be more effective than Hilbert spaces in quantum physics as operators of different cardinality can be considered together.

In rigged Hilbert spaces discrete and continuous bases exist togheter. In particular discrete and
continuous bases coexist such that special functions are transformation matrices between them.

In RHS -consistently with group theory- operators of different cardinality  can be
together generators of a Lie algebra
or elements of an universal enveloping Lie algebra.

The fundamental point is that, while in a HS hermitian operators with spectrum of
different cardinality lead to undefined operations (that, anyway,
physicists are used to perform without too much hesitation), in a rigged Hilbert
space the theory is mathematically consistent.

We have discussed here the complex rigged Hilbert space of quantum mechanics.
The discussion of the real Hilbert space used in signal processing follows exactly
the same lines.


\end{document}